\documentclass[twocolumn,showpacs,preprintnumbers,amsmath,amssymb]{revtex4}
\usepackage{graphicx}
\usepackage{dcolumn}
\usepackage{bm}

\begin{document}
\preprint{JR 04-0020}
\title{Optimization of nanostructured permalloy electrodes \\
for a lateral hybrid spin-valve structure}

\author{T. Last}
\author{S. Hacia}%
\author{M. Wahle}%
\author{S.F. Fischer}%
 \altaffiliation[Electronic mail:]{saskia.fischer@rub.de}
\author{U. Kunze}%
\affiliation{%
Werkstoffe und Nanoelektronik, Ruhr-Universit\"at Bochum, 44780
Bochum, Germany
}%


\begin{abstract}
Ferromagnetic electrodes  of a lateral semiconductor-based
spin-valve structure are designed to provide a maximum of
spin-polarized injection current. A single-domain state in
remanence is a prerequisite obtained by nanostructuring Permalloy
thin film electrodes. Three regimes of aspect ratios $m$ are
identified by room temperature magnetic force microscopy: (i)
high-aspect ratios of $m \ge 20$ provide the favored remanent
single-domain magnetization states, (ii) medium-aspect ratios $m
\sim 3$ to $m \sim 20$ yield highly remanent states with closure
domains and (iii) low-aspect ratios of $m \le 3$ lead to
multi-domain structures. Lateral kinks, introduced to bridge the
gap between micro- and macroscale, disturb the uniform
magnetization of electrodes with high- and medium-aspect ratios.
However, vertical flanks help to maintain a uniformly magnetized
state at the ferromagnet-semiconcuctor contact by domain wall
pinning.

\end{abstract}

\pacs{Valid PACS appear here}
\maketitle
\section{\label{sec:level1}INTRODUCTION}
Lateral ferromagnet-semiconductor (FM-SC) devices have attracted
much interest  for future all-electrical spintronics applications
\cite{DATTADAS90, Prinz95, review, Hacia03} such as spin valves or
spin transistors. Hereby, ferromagnetic electrodes are sought to
spin polarize a current which is subsequently injected into a
semiconductor allowing spin manipulation by purely electrical
means. While other materials such as half-metallics or
ferromagnetic semiconductors are strongly favorable because of
high spin polarization and high spin injection rates they suffer
major drawbacks due to difficult technological processing and the
lack of stable long range ferromagnetism at room temperature.
Until today, only ferromagnetic metals provide high enough Curie
temperatures for reliable room temperature operation.
Unfortunately, ferromagnetic transition metals (Fe, Ni, Co) and
the technological relevant alloys (e.g. Permalloy) only allow  a
moderate spin polarization up to 45~\% \cite{Mod00}. Therefore we
aim at the maximum spin polarization possible by realizing a
uniform magnetization at the FM-SC contact. For zero-field
operation a magnetic single domain state in remanence of the
ferromagnetic electrodes is essential. The magnetization
configuration of soft ferromagnetic thin film elements are
dominantly determined by geometry. This permits tailoring of the
remanent domain configuration and coercive fields by variation of
the length-to-width aspect ratio in rectangular elements. However,
in recent works on lateral FM-SC spin valves
\cite{MON99,Lee99,Hu01,Hu02} multi-domain states at remanence
prevail or magnetic domain states and stray field contributions
were not characterized. A multi-domain configuration of low aspect
ratio Py structures as discussed by G. Meier \textit{et al.}
\cite{Meier1, Meier2} leads to a reduction of the stray field, but
an application for zero-field operation is limited as a maximum
degree of spin-polarization cannot be achieved. Furthermore, for
low-aspect ratio Py platelets a variety of different metastable
remanent magnetic states can be found \cite{AHubert, Hub98,
Hert99}, although a well defined remanent domain state is
favorable \cite{DATTADAS90}.

In this work, we propose a lateral FM-SC spin-valve layout for a
maximum possible spin polarization in remanence (zero-field
operation) and a minimum of well-controlled stray fields.
Single-domain rectangular Py electrodes are tailored by
nanostructuring high-aspect ratios and characterized in detail by
magnetic force microscopy (MFM). For application in a lateral
hybrid device additional lateral kinks and vertical flanks may
have to be considered. Lateral kinks may bridge the gap between
the micro- or nanoscale electrodes and macroscale bond pads, and
vertical flanks may arise from the transition of a thick field
oxide to a thin tunneling oxide at the edge of the FM-SC contact
area. We show that lateral kinks lead to domain wall formation and
vertical flanks act as domain wall pinning centers. Our MFM study
comprises three steps. First, we map the multidomain to
single-domain transition for rectangular Py micro- and
nanostructures over a wide range of aspect ratios. Second, a
reliable maintenance of the highly remanent state in high-aspect
ratio Py wires is proven by application of magnetic fields prior
to imaging.
Third, the influence of lateral kinks and vertical flanks on the
formation of domain walls is studied for a lateral Py-InAs
spin-valve structure.

\section{Preparation and experiment}
In order to map the multi- to single-domain transition rectangular
polycrystalline Py micro- and nanostructures are prepared on p-Si
(111) and undoped n-type InAs (100) by electron-beam lithography
(EBL), electron-beam evaporation (EBV) and lift-off. The length
$l$ varies from 200~nm up to 60~$\mu$m and the width $w$ from
200~nm up to 30~$\mu$m, covering aspect ratios $m = l/w$ from 1 to
105 (Py thickness \textit{t}~=~38~nm and 50~nm).

Lateral Py-oxide-InAs spin-valve structures are prepared by
high-resolution EBL and lift-off patterning. InAs exhibits a
strong spin-orbit interaction which makes it an interesting
candidate for spintronic applications \cite{DATTADAS90, Luo90,
Andrada97}. First, a 200~nm thick electrically isolating SiO$_{2}$
film is sputtered on the InAs substrate. Then a 20~$\mu$m broad
contact area is reopened to the InAs substrate by means of optical
lithography and wet chemical etching. A thin tunneling oxide is
thermally grown at $T$~=~140~$^{\circ}$C in dry O$_{2}$ for 120
min. Finally, the 30~nm thin film Py electrodes are prepared by
patterning a 80~nm PMMA (polymethylmethacrylate) film using EBL at
5~kV, EBV and lift-off in acetone. The electrodes with overall
lengths of 150~$\mu$m are placed across a  20~$\mu$m wide
tunneling contact area. Electrical access to the electrodes is
achieved by bond pads on the thick field oxide outside the FM-SC
contact area. The electrodes 2~$\mu$m and 0.4~$\mu$m wide are
separated by 0.2~$\mu$m.

MFM is performed with a Nanoscope MultiMode scanning-probe
microscope (DI) operating in the
TappingMode/LiftMode$^{\textrm{TM}}$. A commercially available
probe (MESP, DI) is used and pre-magnetized along its vertical
axis. Lift scan heights of about 120 nm are applied to minimize
the probe-induced perturbations.

Two-dimensional micromagnetic simulation were performed by NIST
package OOMMF \cite{oommf}.

\section{RESULTS AND DISCUSSION}
\subsection{\label{sec:level2}Multidomain to single-domain transition}
Fabrication of single-domain ferromagnetic electrodes for a
zero-field operation in FM-SC hybrid devices requires a precise
knowledge of the influence of electrode geometry on remanent
magnetization. The remanent domain configuration of a rectangular
thin film Py structure can accurately be controlled by its aspect
ratio $m$ because the magnetocrystalline anisotropy is negligible
at room temperature and the resulting magnetization arrangement is
determined by the competition between the exchange energy and the
stray field energy. Here, we map the multi- to single domain
transition for the range of aspect ratios $m$ from 1 to 70. We
find, in accordance with Gomez \textit{et al.} \cite{GOMEZ99},
that Py microstructures of low-aspect ratio (\textit{m} from 1 to
3) invoke five principle multidomain patterns. As an example, the
multidomain MFM image of a low-aspect ratio ($m$ = 2.5)
rectangular Py structure is shown in Fig.~\ref{plot} (a),
micrograph (I). Dark and light contrasts indicate perpendicular
stray field components of the tilted diamond pattern consisting of
seven domains in flux closure configuration.  Aspect ratios
\textit{m} of about 3 denote a transition from multidomain states
to highly remanent magnetization states which originate for $m \ge
$~4. MFM images of highly remanent domain states consist of a
featureless central region which indicates a uniform magnetization
parallel to the long axis of the microstructures and dark or light
stray field contrast from fan-type closure domains at both ends.
The extension of the closure domains diminish and the regions of
homogeneous magnetization increase with increasing aspect ratio.
An example of a MFM image exhibiting highly remanent magnetization
states is given for two Py stripes with medium aspect ratios of
$m$~$\approx$~14 in Fig.~\ref{plot} (a), micrograph (II). Finally,
single domain states are observed for high-aspect ratios of $m \ge
20$. By MFM imaging closure domains are not anymore resolved and
perpendicular components of stray fields at the wire ends are
marked as structureless black and white spots as shown in
Fig.~\ref{plot} (a), micrograph (III) for two Py stripes with
high-aspect ratios of $m$ = 57 (top) and $m$ = 43 (bottom).
\\
The transition from a multi- to a single-domain state is
summarized in Fig.~\ref{plot} (b). The plot establishes validity
regions of (I) multidomain patterns, (II) highly remanent
magnetization states and (III) single domain states for the
investigated structure sizes $l$ and corresponding aspect ratios
$m$. A transition line from single-domain state to highly remanent
state is included from a phenomenological model developed by A.
Aharoni \cite{AHA88} for ellipsoidal particles. Below a lower
bound, namely the minor semiaxis a$_{c0}$, ellipsoidal particles
are found to be in a single-domain state in remanence. The lower
bound depends on the exchange interaction, on the saturation
magnetization, and on the aspect ratio between the particle$'$s
major and minor axis $m'$. According to Seynaeve \textit{et al.}
\cite{SEY01} a critical length $L_{c} = 2m'a_{c0}$ separates the
multidomain states (above $L_{c}$) from the single-domain states
(below $L_{c}$). For the calculation of $L_{c}$ the bulk
saturation magnetization of Py, 800~kA/m, is taken and an exchange
constant of 1.3$\cdot10^{-9}$~J/m is required, 2 orders of
magnitude higher than the reported bulk value of Py \cite{Hert99}.
A comparable deviation was found in the analogue evaluation of
domain structures for Co microstructures \cite{SEY01}. While the
model is based on ellipsoidal particles, we investigate
rectangular Py microstructures, and this profoundly changes the
calculation of the stray field. Furthermore, the prepared
structures exhibit finite surface and edge roughness due to vapor
deposition and lift-off, which introduces magnetization
inhomogeneities at the edges. Finally, experimental MFM data
exhibit a resolution limit of the order of the probe lift-scan
height, which is approximately 120 nm. Therefore, the
phenomenological model \cite{AHA88} cannot be expected to match
quantitatively, however, the observed qualitative agreement of the
multi- to single-domain transition indicates the same underlying
fundamentals of scaling behavior.

\subsection{\label{sec:level3}Lateral spin-valve configuration in remanence}
A lateral FM-SC hybrid spin valve consist of two spin-polarizing
and -detecting ferromagnetic electrodes placed on top of a
semiconductor substrate (spin transport channel). For a spin-valve
effect distinctly different switching fields of the ferromagnetic
electrodes are required. As shown above, a single domain state in
remanence is provided by elongated rectangular Py electrodes of
high aspect ratios. The switching field of elongated Py wires are
essentially determined by their widths and our magnetotransport
studies on single Py wires show that the switching field increases
linearly with reciprocal wire width \cite{Last03}.

Below, we demonstrate two prerequisites for  magnetotransport
experiments in order to resolve subtle magnetoresistance effects
on behalf of spin injection. First, a reproducible, stable
remanent magnetic domain configuration after magnetic field sweeps
is required. Second, side effects as stray field contributions
have to be kept to a minimum.

Prior to MFM imaging Py wires of aspect ratios 14 to 33 were
magnetized by applying longitudinal magnetic fields of different
amplitude and polarity. MFM images of single domain and highly
remanent Py wires are shown in Fig.~\ref{arrayhys}. Wire lengths
of 20 $\mu$m resemble the electrode sections within the FM-SC
contact area in the lateral FM-SC spin-valve structure discussed
in section D. The two wires labelled as (I) and (II) exhibit
typical widths for lateral spin-valve devices of 0.5~$\mu$m and
1.4~$\mu$m, respectively. After applying a magnetic field of +60
mT (Fig.~\ref{arrayhys} (a)) all wires are magnetically saturated.
In a reversed field  of -20 mT (Fig.~\ref{arrayhys} (b)) the broad
wire II is switched. At fields of -60 mT all wire magnetizations
are reversed. This subsequent switching occurs all the same in a
reversed magnetic-field sweep. Hence, for parallel Py wires of
different aspect ratios magnetic field induced parallel and
antiparallel magnetization configuration are stable in remanence.
Furthermore, for wires such as (I) and (II) an antiparallel
magnetization configuration exists over a field range of several
ten mT which is suitable for spin blockade experiments.

The lateral extension $\Delta l$ over which perpendicular stray
fields arise from magnetic inhomogeneities at the wire ends can be
roughly quantified from MFM images. Due to the finite radius of
the MFM tip as well as the distinct scan height the ratio of
$\Delta l/l$ of Fig.~\ref{arrayhys} gives an upper bound, which is
given in Fig.~\ref{closdom}. As a lower limit, we have also
inserted in Fig.~\ref{closdom} the values of the lateral extension
of magnetic inhomogeneities in simulated two-dimensional
structures with corresponding aspect ratios but lower total length
(2 $\mu$m). For higher aspect ratios $\Delta l$ is clearly
reduced. It is noteworthy that the electrodes of the lateral
spin-valve structure proposed in section~D have much larger aspect
ratios of 375 and 75. The FM-SC contact area exhibits no wire
ends. Hence, the lateral extension of magnetic inhomogeneities at
the edges of the FM-SC contact area is expected to be strongly
reduced or even nil.

\subsection{\label{sec:level4} Lateral kinks in medium aspect ratio wires}
The influence of lateral kinks on the highly remanent domain
configuration of elongated Py microstructures is outlined for
medium aspect ratios $m$ $\geq$ 4. The domain configuration is
strongly perturbed by kink angles between 10$^{\circ}$ and
90$^{\circ}$, as can be seen from MFM images of Py wires with an
aspect ratio  of 10  ($w$~=~2~$\mu$m, $l$~=~10~$\mu$m) in
Fig.~\ref{kink1}. An undisturbed, straight wire (Fig.~\ref{kink1}
(a)) features the highly remanent domain configuration with
significant closure domains at both ends and a uniformly
magnetized mid-part. However, a 10$^{\circ}$ kink changes the
domain configuration drastically (Fig.~\ref{kink1} (b)). Here,
cross-tie walls separate two main domains. With increasing kink
angle the domain configuration transforms back to highly remanent
states, but a significant magnetization inhomogeneity at the
centre of the structure is observable (Fig.~\ref{kink1} (c), (d)).
For kink angles larger than 70$^{\circ}$ domain walls are clearly
visible at the kink. Micromagnetic two-dimensional simulation of
the magnetization distribution of similar elongated Py wires with
aspect ratio $m$~=~10 show equally that small lateral kink angles
of 10 degree suffice to introduce magnetic inhomogeneity
(Fig.~\ref{simsel}). For the simulation the structure length had
to be reduced to 2~$\mu$m, and a discrete cell size of 10~nm were
chosen.

In summary, horizontal kink angles introduced in a Py wire of
medium aspect ratio disturb the highly remanent domain state.
Fortunately, as shown in section~D vertical flanks prevent the
originated magnetic inhomogeneities from extending  into the FM-SC
contact of a lateral spin-valve.

\subsection{\label{sec:level5}Lateral kinks and vertical flanks in
high aspect ratio wires: optimized spin-valve geometry}

Following the above domain investigations an optimized  electrode
geometry of a lateral spin-valve structure can be set up to study
spin-dependent transport phenomena in low- or zero-field
operation. In order to unambiguously identify electrical
spin-injection in magnetotransport  the lateral spin-valve
structure must fulfill the following requirements: stable
single-domain states in remanence at the FM-SC contact, distinctly
different switching fields of two electrodes, minimized
perpendicular stray field components and multi-terminal
measurement geometries to identify all magentoresistance
contributions.

A top view of a complete device structure is given  by the
scanning electron micrograph (SEM) in Fig.~\ref{rem}. Two broad
and one narrow high-aspect ratio Py electrodes of 30~nm thickness
are spaced by 200~nm in the FM-SC contact area. The inset of
Fig.~\ref{rem} enlarges the  high-quality nanostructured Py
electrodes. For the broad electrodes lateral kinks lead to
macroscopic bond pads (not shown). As the narrow electrode
dominates the magnetoresistance, it is kept straight in order to
avoid any magnetization inhomogeneity in the FM-SC contact area.
All three electrodes approach the FM-SC contact window via
inclined planes of the etched 200 nm thick sputtered SiO$_{2}$
which are depicted in the following as vertical flanks. A
schematic cross-section of the device is shown in Fig.~\ref{sio2}.
An upper bound for the inclination angle of 30 to 40 degree is
found from a cross-sectional SEM of thermally grown and
subsequentially wet-etched SiO$_2$, giving an upper estimate of
the lateral width of the flank of about 100 nm.

Due to the overall length of 150~$\mu$m and the height difference
of 200~nm the complete electrodes structure cannot be captured in
one single MFM image. However, decisive parts of the domain
configuration in the Py electrodes are detailed in MFM images in
Fig.~\ref{svconfmfm}. The narrow electrode has an overall aspect
ratio of 375 ($l$ = 150 $\mu$m, $w$ = 0.4 $\mu$m) and solely black
and white contrasts (Fig.~\ref{svconfmfm} (4),(5)) at the ends of
the wire are visible. The steep flanks which lead to the 20 $\mu$m
long FM-SC contact area do not introduce domain walls in this
straight high-aspect ratio electrode. The narrow electrode is in a
single-domain state in remanence. In consequence, vertical flanks
do not impose a multidomain state with domain walls in straight
high-aspect ratio wires.

Contrary, the domain configuration of high-aspect ratio electrodes
with lateral kinks is more complex. The broad electrodes with an
overall aspect ratio of 73 ($l$ = 145 $\mu$m, $w$ = 2 $\mu$m) are
composed of several parts: the 45~$\mu$m long wire ends ($m$=21)
which are connected to a straight mid-part by a lateral kink angle
of 45 degree. The mid-part consists of three 20~$\mu$m long
straight wire sections ($m$~=~10). MFM images of these wire
sections are depicted in Fig.~\ref{svconfmfm} (b) by images 1, 2,
5 and 6, which show a uniform magnetization parallel to the length
of the wire sections and strong stray field contrasts at the wire
ends. At the lateral kink either a magnetization inhomogeneity
(Fig.~\ref{svconfmfm} (b) 1, 5) or a domain wall occurs
(Fig.~\ref{svconfmfm} (b) 2, 6).  However, at the edge of the
FM-SC contact area the vertical flanks pin domain walls so that
also the broad electrodes are uniformly magnetized within the
region of the FM-SC contact. A MFM image of the decisive section
of the FM-SC contact area is displayed in Fig.~\ref{svconfmfm}
(c), 10, showing uniform remanent magnetization for all three
electrodes. From the topography, Fig.~\ref{svconfmfm} (c), image
11, it is evident that the lift-off edges with a roughness less
than the wire thickness ($\leq$~30 nm) invoke minimal magnetic
contrasts due to fringe fields, as visible in Fig.~\ref{svconfmfm}
(c), image 10.

In magnetotransport experiments perpendicular stray field
components, visible by dark and light contrast in the MFM images,
may induce local Hall voltages in the semiconductor \cite{MON99}.
Such side effects complicate the study of spin-dependent transport
considerably. However, in the lateral spin-valve structure
presented, stray field contributions are minimized and
well-controllable. First, strong perpendicular stray fields occur
only at the vertical flanks because the electrodes´ sections in
the  FM-SC contact area are in a single-domain state. Furthermore,
the remaining stray fields occur at the vertical flanks for which
a tunneling injection current is exponentially reduced with
increasing oxide thickness. Any remaining contribution of fringe
fields to the magnetoresistance can be precisely characterized and
subtracted by complementary four-terminal measurements.

\section{\label{sec:level6}Conclusion}
A lateral ferromagnet-semiconductor device structure is optimized
for spin-valve operation by nanostructuring  high-aspect ratio
(\textit{m} $\gtrsim$ 20) Py electrodes. A combination of lateral
kinks and vertical flanks necessary in hybrid devices provides the
favored single-domain state of the spin-injection and -detection
electrodes in remanence. The advantages of the presented lateral
spin-valve layout  will apply similarly to future spin-polarizing
materials that exhibit long range ferromagnetism.

\begin{acknowledgments}
T. Last gratefully acknowledges financial support by the
Evangelisches Studienwerk Haus Villigst e.V. This work has been
supported by the Deutsche Forschungsgemeinschaft within the
Sonderforschungsbereich 491.
\end{acknowledgments}

\newpage 
\renewcommand{\baselinestretch}{1.2}

\newpage

\begin{figure}[htb]
\includegraphics[scale=.9]{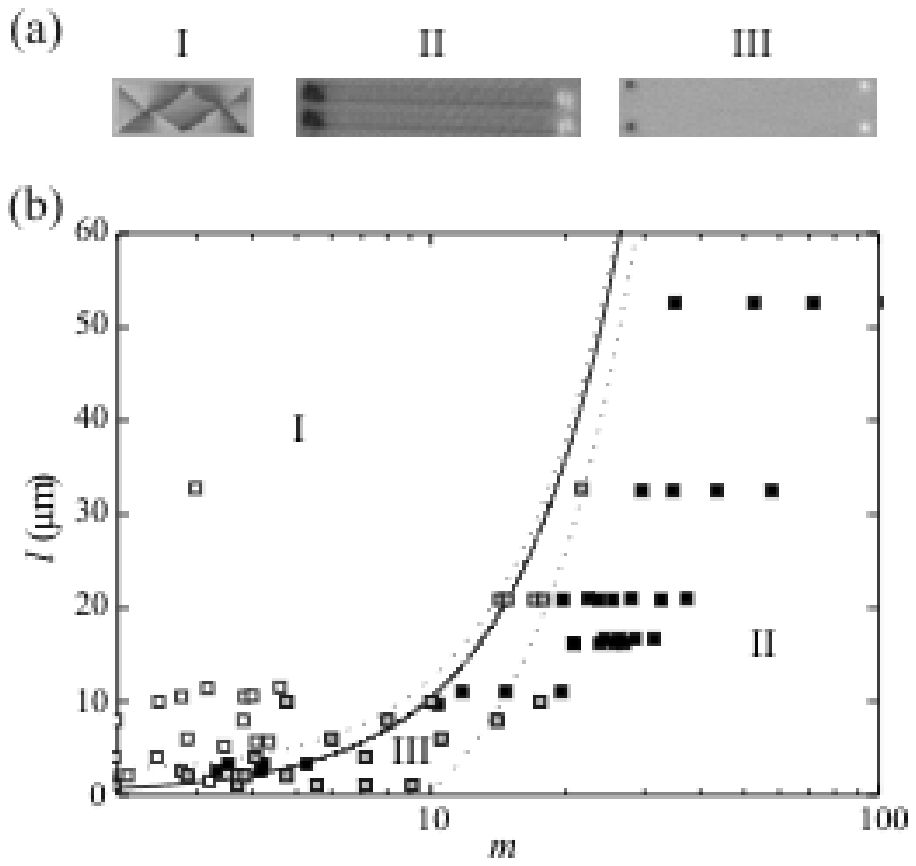}
\vspace{4.0cm}

\caption{\label{plot}}
\end{figure}

\begin{figure}[hbt]
\includegraphics[scale=.9]{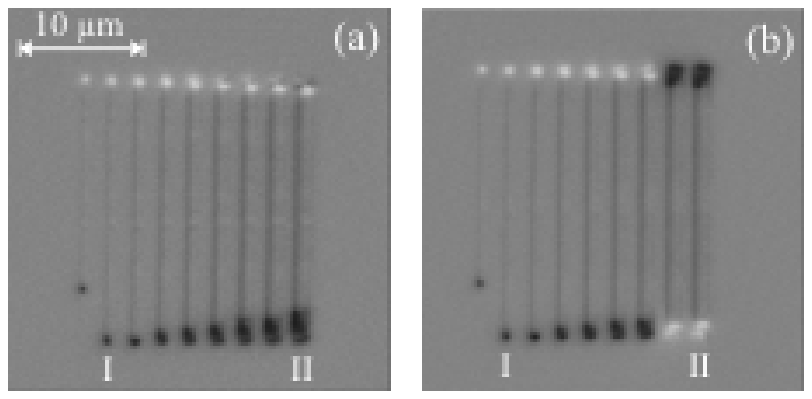}
\vspace{4.0cm}

\caption{\label{arrayhys}}
\end{figure}

\begin{figure}[h]
\includegraphics[scale=.9]{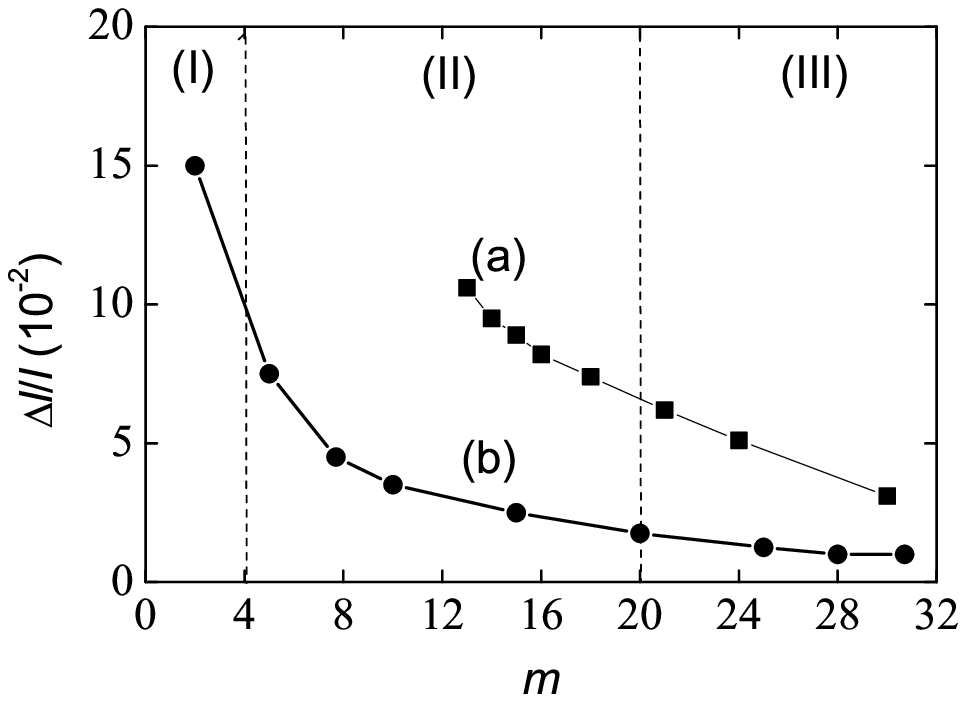}
\vspace{4.0cm}

\caption{\label{closdom}}
\end{figure}

\begin{figure}[h]
\includegraphics[scale=.5]{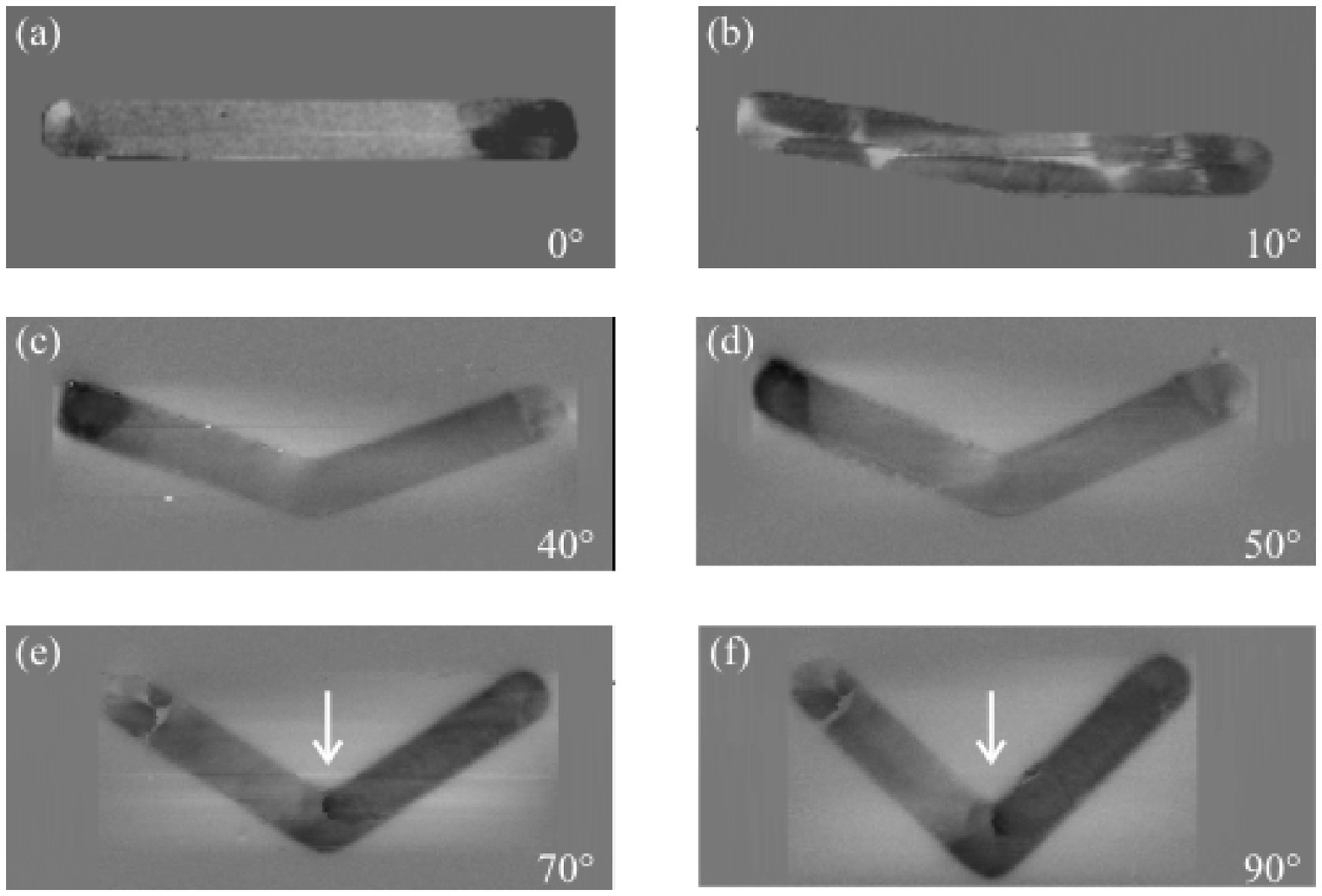}
\vspace{4.0cm}

\caption{\label{kink1}}
\end{figure}

\begin{figure}[h]
\includegraphics[scale=.9]{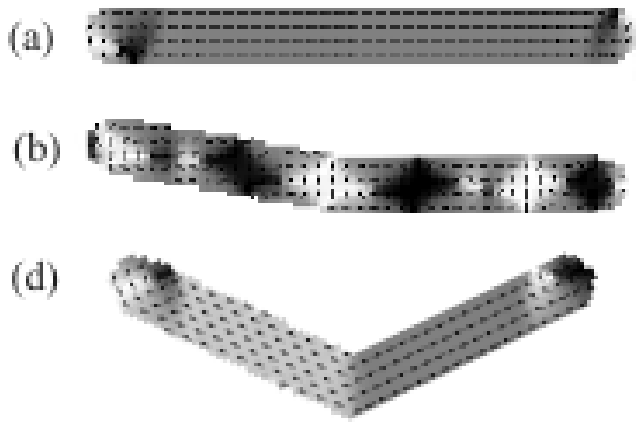}
\vspace{4.0cm}

\caption{\label{simsel}}
\end{figure}

\begin{figure}[h]
\includegraphics[scale=.5]{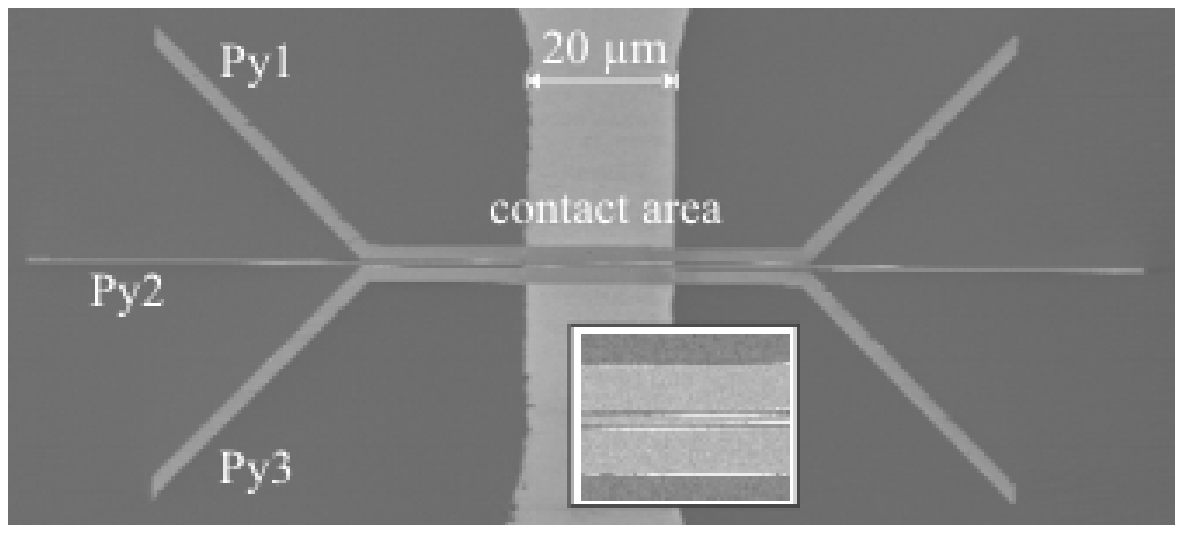}
\vspace{4.0cm}

\caption{\label{rem}}
\end{figure}

\begin{figure}[h]
\includegraphics[scale=.5]{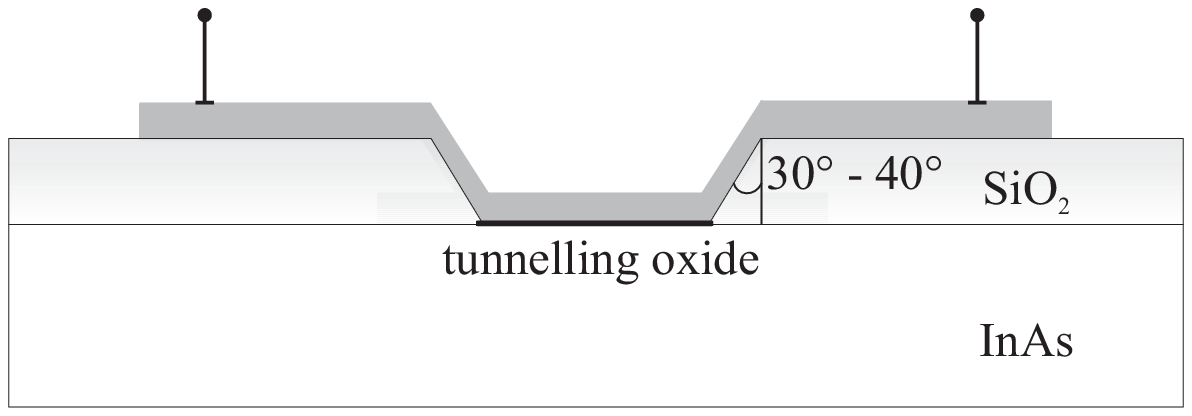}
\vspace{4.0cm}

\caption{\label{sio2}}
\end{figure}

\begin{figure}[h]
\includegraphics[scale=.5]{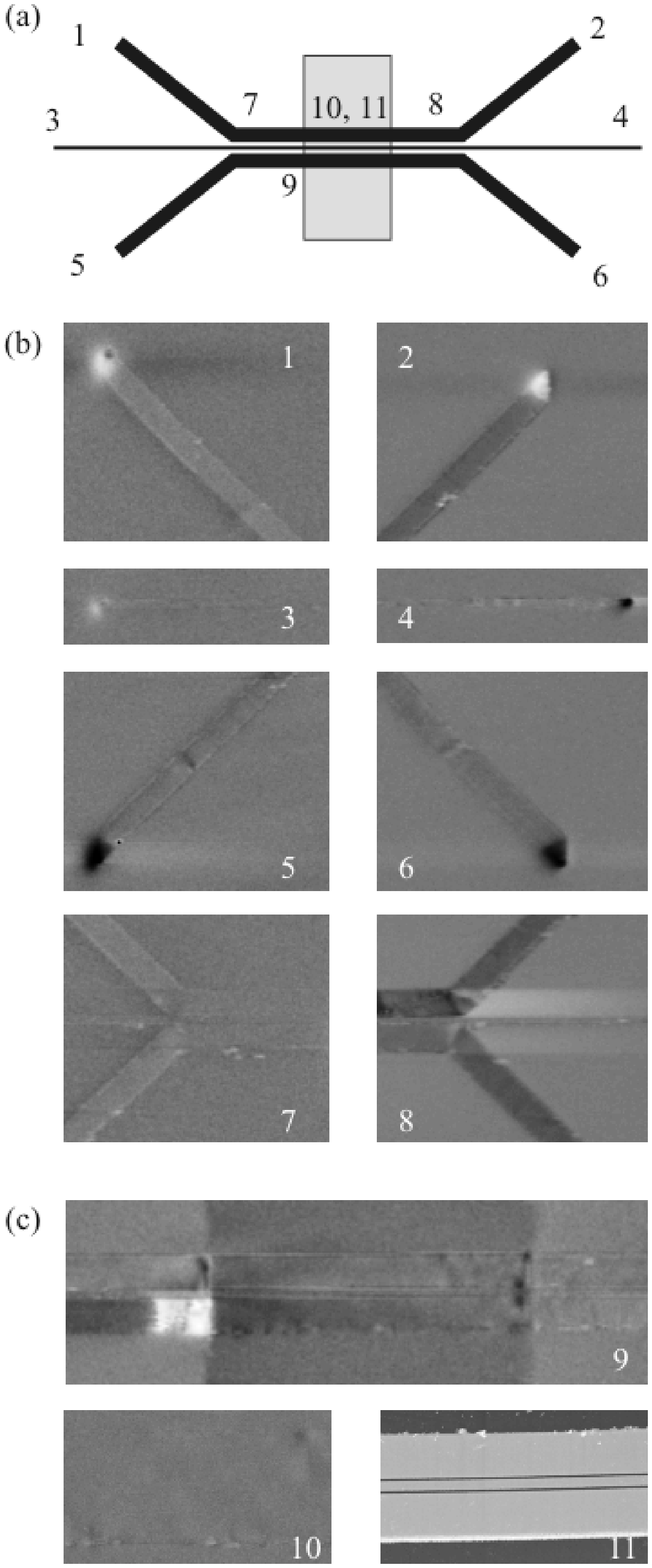}
\caption{\label{svconfmfm} }
\end{figure}

Figure~\ref{plot}: { (a) Magnetic force microscopy images of Py
structures in remanence with (I) low ($m$ = 2.5, with $l$ = 10
$\mu$m, \textit{t}~=~38~nm), (II) medium ($m$ $\approx$ 14, $l$ =
20.9 $\mu$m, \textit{t}~=~38~nm) and (III) high aspect ratios ($m$
= 57 (top wire), $m$ = 43 (bottom wire), \textit{t}~=~50~nm). (b)
Experimentally observed domain states of Py micro- and
nanostructures for varied length \textit{l} versus aspect ratios
\textit{m}. Open squares indicate multidomain states (I), grey
squares denote high remanent states (II) and closed squares
indicate a single-domain state (III). The black solid line depicts
the transition line between multi- and single-domain configuration
from a phenomenological model \cite{AHA88}. The dotted lines are
guides to the eye.}
\\\\
Figure~\ref{arrayhys}: {MFM images of an array of Py wires
($l$~=~16.7~$\mu$m, 20.9~$\mu$m) with aspect ratios from 14.2 to
32.7. Reliable maintenance of highly remanent state after applying
longitudinal magnetic fields of: (a) +60~mT and (b) -20~mT are
observed.}
\\\\
Figure~\ref{closdom}: {Ratio of the lateral extension of magnetic
inhomogeneities to the structure length $\Delta l/l$ versus the
aspect ratio $m$ of Py wires. (a) Upper estimates of $\Delta l/l$
from MFM images in \ref{arrayhys} of an array of Py wires with
medium and high-aspect ratios. (b) Lower limit for $\Delta l/l$
from two-dimensional micromagnetic simulation (wire length: 2
$\mu$m). Aspect ratio regions are denoted as (I) low, (II) medium
and (III) high.}
\\\\
Figure~\ref{kink1}: {MFM images of medium-aspect ratio Py wires
($w$~=~2~$\mu$m, length, $l$~=~10~$\mu$m, $m$ = 10) with lateral
kinks of (a) 0$^{\circ}$, (b) 10$^{\circ}$, (c) 40$^{\circ}$, (d)
50$^{\circ}$, (e) 70$^{\circ}$ and 90$^{\circ}$ (f). Arrows in (e)
and (f) depict the insertion of domain walls.}
\\\\
Figure~\ref{simsel}: {Micromagnetic two-dimensional simulation of
the magnetization distribution in elongated Py wires of medium
aspect ratios ($m$~=~10) with lateral kinks (a) 0$^{\circ}$, (b)
10$^{\circ}$, (c) 50$^{\circ}$. Structure length: 2 $\mu$m, cell
size: 10 nm.}
\\\\
Figure~\ref{rem}: {Scanning-electron-microscope image of a lateral
Py-oxide-InAs spin-valve structure with electrodes of widths 2
$\mu$m and 0.4 $\mu$m separated by 0.2 $\mu$m in the contact area
(electrode length: 150 $\mu$m). Inset: Enlargement of the Py
electrodes at the FM-SC contact.}
\\\\
Figure~\ref{sio2}: {Schematic cross-section of the lateral
spin-valve structure. The etched vertical SiO$_{2}$ flanks at the
edges of the ferromagnet-semiconductor contact area have
inclination angles of 30$^{\circ}$ to 40$^{\circ}$.}
\\\\
Figure~\ref{svconfmfm}: {MFM images of a lateral Py-oxide-InAs
spin-valve configuration. High-aspect ratio Py electrodes of 50 nm
thickness are separated by 200 nm in the region of the contact
area. (a) Schematic view of the complete structure, 150 $\mu$m in
length, with section numbering of MFM images displayed in (b) and
(c).(b) Stray field contrast at the ends of the broad wires
(images 1,2,5,6) result from a multidomain configuration which
originates from lateral kinks (images 7 and 8). The pair of strong
black/white contrasts at the closures of the straight narrow wire
(images 3 and 4) indicates a single domain magnetization
configuration.}
\\

\newpage
\large{T. Last \emph{et al.}, Figure 1}
\vspace{4.0cm}

\newpage
\large{T. Last \emph{et al.}, Figure 2} \vspace{4.0cm}

\newpage
\large{T. Last \emph{et al.}, Figure 3} \vspace{4.0cm}

\newpage
\large{T. Last \emph{et al.}, Figure 4} \vspace{4.0cm}

\newpage
\large{T. Last \emph{et al.}, Figure 5} \vspace{4.0cm}

\newpage
\large{T. Last \emph{et al.}, Figure 6} \vspace{4.0cm}

\newpage
\large{T. Last \emph{et al.}, Figure 7} \vspace{4.0cm}

\newpage
\large{T. Last \emph{et al.}, Figure 8} \vspace{2.0cm}

\end{document}